# Improving Image Search based on User Created Communities

Amruta Joshi*   Junghoo Cho†   Dragomir Radev ‡

Ahmed Hassan §

October 11, 2018

## Abstract

Tag-based retrieval of multimedia content is a difficult problem, not only because of the shorter length of tags associated with images and videos, but also due to *mismatch in the terminologies used by searcher and content creator.* To alleviate this problem, we propose a simple concept-driven probabilistic model for improving text-based rich-media search. While our approach is similar to existing topic-based retrieval and cluster-based language modeling work, there are two important differences: (1) our proposed model considers not only the query-generation likelihood from cluster, but explicitly accounts for the overall "popularity" of the cluster or underlying concept, and (2) we explore the possibility of inferring the likely concept relevant to a rich-media content through the user-created communities that the content belongs to.

We implement two methods of concept extraction: a traditional cluster based approach, and the proposed community based approach. We evaluate these two techniques for how effectively they capture the intended meaning of a term from the content creator and searcher, and their overall value in improving image search. Our results show that concept-driven search, though simple, clearly outperforms plain search. Among the two techniques for concept-driven search, community-based approach is more successful, as the concepts generated from user communities are found to be more intuitive and appealing.

*Department of Computer Science, University of California, Los Angeles, CA, USA; Email: amrutaj@cs.ucla.edu

†Department of Computer Science, University of California, Los Angeles, CA, USA; Email: cho@cs.ucla.edu

‡Department of Computer Science, University of Michigan, Ann Arbor, MI, USA; Email: radev@umich.edu

§Department of Computer Science, University of Michigan, Ann Arbor, MI, USA; Email: hassanam@umich.edu

# 1 Introduction

With the rise in the number of "Web 2.0" sites in recent years, more and more users are sharing their own rich-media content online. From being mere read-only consumers of web data, users now actively contribute rich multimedia content to the web. To help users explore and access the user-created rich-media content, it is critical to provide effective search services on such content. Unfortunately, the current-generation search engines on rich-media content often fail to return what the users expect for many queries. For example, when a searcher types the query 'Jasmine' to a photo sharing website like Flickr, as seen in Figure 1, typical results include a random girl's picture with the name 'Jasmine', a dog with the name 'Jasmine' and so on. Similar problems are seen in search results of other photo sharing websites like Google Picasa Community Photo Search [1] and SmugMug Photo Sharing [2].

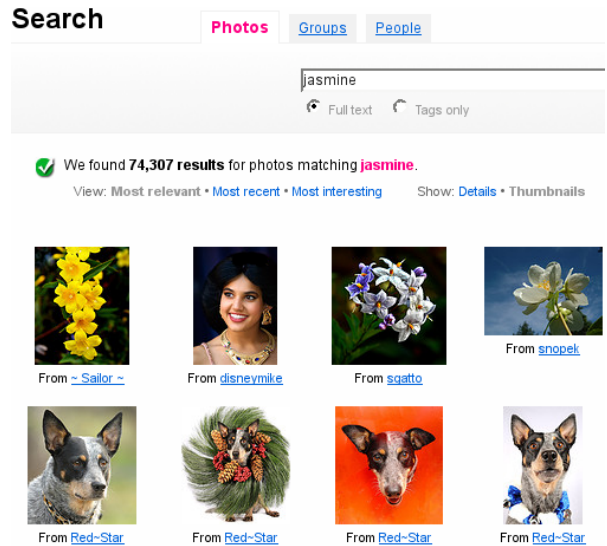

Figure 1: Search Results for query "jasmine" on Flickr

We observe that the poor quality of search results on user-created rich-media content is often due to the inherent *mismatch of the "meaning" or "concept" that the content creator and the searcher associate with a keyword.* For instance, in case of the jasmine example, the tag 'Jasmine' for a dog's image may be appropriate for the creator, because the dog's name 'Jasmine' may be the most important word that distinguishes it from other images uploaded by the creator. In contrast, when a searcher issues the query 'Jasmine' to a search engine, he/she is likely to look for an image in its "popular" meaning, like jasmine the flower or jasmine the tea. Unfortunately, existing search engines mainly rely on the syntactic similarity between the content keywords and the query, so they typically consider a dog's image as relevant to the query 'Jasmine' as a flower's image as long as both of them have the keyword 'Jasmine' associated with them.

---

[1] http://picasaweb.google.com
[2] http://www.smugmug.com

To address this problem, we propose that we should explicitly identify and employ the underlying concepts of media content during search. That is, instead of computing the keyword-based similarity directly between an image and the query, we first identify the underlying *concept* that the image is associated with, compute the likelihood that the concept is *what the searcher is looking for*, and return the images *sorted by this likelihood*. The hope of this concept-driven search is that, this way, the images corresponding to the popular concepts related to the query 'Jasmine' are returned at the top, not random images that happen to be associated with the keyword 'Jasmine'.

A large body of prior work tries to address the concept mismatch problem in the context of textual document retrieval (for example, see topic detection [5, 3], cluster-based retrieval [8] and cluster-based language modeling [13] work). While this thread of work, in particular the cluster-based language modeling, is very close to our approach, a few key differences make the search on user-created rich-media content unique and more challenging. First, the methods used for automatic topic (or cluster) detection often lead to topics that are difficult to interpret by end users due to the inherent limit of automatic methods. Second, in existing work, the ranking of topics are mainly based on the *relevance* of each topic to the query and often fails to rank the "most likely" meaning the searcher has in mind at the top. Third, the language used to tag user-created rich media is often very informal and inconsistent. For instance, many users on Flicker tag the image of their cars as "my baby", meaning that the cars are dear to their heart, not that they are baby pictures. These factors make the direct application of prior work to the rich-media search not as effective as they are when applied to a textual-document collection.

In order to address these problems, our concept-driven search approach exploits our two observations on user-created rich-media content: First, users tend to organize and share their content through self-elected communities of their interest, such as the community of "Nature lovers" or the community of "Baby pictures". As we will explain in more detail later, communities provide strong hints on the likely concept that the content creator had in mind when he used a particular keyword as its tag. Second, when a searcher uses a particular keyword in his search query, the likely meaning of the keyword is often the "most popular" interpretation of the keyword. Based on these observations, in our concept-driven search, we investigate using the communities that a content belongs to in order to identify the concept of a particular content is associated with. We also explicitly model and incorporate the "popularity" of each concept as an essential component of ranking when we consider each concept as potential match.

In summary, we make the following contributions in this paper: First, we present a formal model for concept-driven search, explicitly modeling the underlying concept through a probabilistic approach incorporating the notion of popularity. Further, we investigate how we can use the commu-nity within social web sites to identify the likely concepts of the contents *and* queries. Finally, we conduct extensive experiments to evaluate the effectiveness of our approach. Our experiments show that communities are indeed great sources for identifying the likely concepts of media content and the queries.

## 2 Concept-Driven Search

While we believe our approach can be applied to general user-created rich-media content on the web, in this paper, we mainly focus on the image search as our driving application.

Performing concept-driven search on images requires us to develop solutions to the following three tasks.

*1. Capture the underlying concepts of the images.* For example, if the image description says 'jasmine' we want some way of determining whether the image is about a flower or a girl or a pet.

*2. Capture the searcher's expected concept.* While the expected concept may be different for each searcher, we conjecture that a general searcher is more likely to look for the popular concepts related to the search keyword[3]; for instance, flower is probably the more popular expected concept for the keyword 'jasmine' than a dog's name. This suggests that in identifying the searcher's expected concept, we need to measure both (1) the *relevance* of a concept to the query and (2) the general *popularity* of the concept; a concept is likely to be what is being searched for, if it is both relevant to the query *and* popular.

*3. Identify appropriate images for the matched concepts.* Once we have determined the concepts that are of interest to the searcher, we need a mechanism to pick out the right images matching that concept to serve search results.

More formally, concept-driven search can be stated and understood using a probabilistic model. Let $Q$ be a query and $I$ be an image. Query $Q$ can be expressed as a term vector $Q = (t_1, t_2, ..., t_n)$, where $t_1, t_2, ...$ are the individual terms (and their associated term-and-inverse-document frequencies) in the query. Image $I$ can also be expressed as a term vector $I = (t_1, t_2, ..., t_n)$, where $t_1, t_2, ...$ are the terms associated with the image as text descriptors (or tags). Then, the common way of measuring the relevance of $Q$ to $I$ is computing the cosine similarity between their term vectors. In our work, however, we assume that the relevance of $Q$ to $I$ is indirectly measured through the relevance of the concepts of the query $Q$ and the image $I$. That is, assuming that $C_j$'s represent possible concepts of the query $Q$, we compute the relevance of $I$ to $Q$ by summing up the relevance of the two through $C_j$'s:

$$P(I|Q) = \sum_{C_j} P(I, C_j|Q)$$

---

[3]In this paper, we limit ourselves to a general searcher's perspective. Another possible approach is to capture the searcher's perspective better by personalization. We leave this possibility as future work

Using Bayes rule, this can be restated as

$$P(I|Q) = \sum_{C_j} P(C_j|Q)P(I|C_jQ). \qquad (1)$$

Here, the first term $P(C_j|Q)$ can be further restated using Bayes rule:

$$\begin{aligned} P(C_j|Q) &= \frac{P(C_jQ)}{P(Q)} \\ &= \frac{P(Q|C_j)P(C_j)}{P(Q)}. \end{aligned} \qquad (2)$$

Combining equations (1) and (2), we get

$$P(I|Q) = \frac{1}{P(Q)} \sum_{C_j} P(Q|C_j)P(C_j)P(I|C_jQ). \qquad (3)$$

Note that the individual factors on the right-hand side of the equation express how we can use concepts to score images. First, the factor $1/P(Q)$ is independent of the images and can be ignored for ranking because it does not affect the relative ranking of images. The term $P(Q|C_j)$ is the likelihood of $Q$ given $C_j$ and can be interpreted as the relevance of the query $Q$ to the concept $C_j$. The next term $P(C_j)$ is the probability of the concept $C_j$ and can be interpreted as the general "popularity" of the concept $C_j$. The last term $P(I|C_jQ)$ is the likelihood of $I$ given $Q$ and $C_j$, and can be interpreted as the relevance of the image $I$ within the concept $C_j$ of the query $Q$. Interestingly, we observe that this probabilistic model of the concept-driven scoring leads to both (1) the query-independent popularity score of the concept $C_j$ and (2) the relevance score of the concept $C_j$ to the query $Q$ in computing the score that we discussed in the beginning of this section. Using this formal model, we study how we can obtain the concept $C_j$, and how we can compute the individual scoring functions, $P(Q|C_j)$, $P(C_j)$, and $P(I|C_jQ)$, given a concept, a query, and an image.

## 3  Extracting Concepts

We explore two approaches to extracting concepts for images: community-based approach, and cluster-based approach. In both cases, we represent the extracted concepts using concept tags, which are representative labels of the community or cluster from which the concept was extracted.

### 3.1  Community-Based Concept Extraction

A community on the web is a group of users coming together to share information about a common topic of interest. For example, On Flickr, each community has a pool where members share their images related to the community. Flickr has more than 1 million communities on topics like "Nature exploration", "Baby pictures", "Fruit lovers", etc. If the users of each community have an image relevant to their shared interest, they explicitly add it to be part of the community.[4] Therefore, an image belonging to a community on "Baby pictures" is likely to be a picture related to a baby than anything else, which suggests that the community membership of an image is a strong indication of the possible concept an image is associated with. As communities contain a collection of images from a number of users, their manually-tagged aggregated knowledge on provides a better concept identity for each image, which may not be otherwise captured by stand-alone tags.

To leverage communitites as concepts, we need a mechanism to compute $P(C_j)$, $P(Q|C_j)$, and $P(I|C_jQ)$ from equation (3) using communities. For the reader's convenience, we show the original concept-driven search equation here again.

$$P(I|Q) = 1/P(Q) \sum_{C_j} P(Q|C_j)P(C_j)P(I|C_jQ)$$

### 3.1.1 Query Independent Concept Score $P(C_j)$.

The query-independent score of a concept, represented as $P(C_j)$ in equation (3), is the "popularity" value of the community irrespective of the query. This can measured by a number of different community-attributes such as, the number of members in the community, the number of shared objects, the amount of activity on discussion threads, etc. In our evaluation, for simplicity and ease of implementation, we use the the log of the number of members in the community as its query-independent popularity score. Even this simple measure proves to be very useful in capturing community popularity.

### 3.1.2 Community Representation

Since our objective behind using communities is to use the collective knowledge of members related to the specific topic of interest, we represent a community as an aggregate of all tags associated with the images shared within the community. More formally, the textual representation of a community $Comm$ is the tag frequency vector $Comm = (T_1, T_2, ..., T_n)$, where $T_1, T_2, ...$ are aggregated tag counts from individual image vectors for images shared in the community

$$Comm(T_1, T_2, ..., T_n) = \sum_{I \epsilon Comm} I(t_1, t_2, ..., t_n)$$

We then eliminate the very low frequency terms from the community vector (i.e., terms with frequency below 2 standard deviations from the mean term frequency) to reduce the effect of outliers. Further, we normalize the community vector such that the $i^{th}$ element of the $Comm$ vector approximately represents the probability of term $t_i$ in that community.

---

[4] An image may be added to multiple communities on Flickr if the user finds his image relevant to different communities.

### 3.1.3 Concept Relevance for Query $P(Q|C_j)$.

While representing concepts using communities, we may come across mutiple communitites about the same concept. In this case, we need to merge these to a single concept that they represent. Therefore, we cluster highly similar communities with ($> 90\%$) similarity among their normalized term vectors $Comm_i$ into a single concept. We use simple k-means clustering algorithm to accomplish this. Thus, we represent a concept $C_j$ as a weighted aggregate of $n$ community vectors $Comm_1, Comm_2, ..., Comm_n$ in the concept cluster, with community weight proportional to the number of images contained in it.

As community clustering and concept vector computation are completed during the offline process of indexing, the query-time performance of our system remains unaffected. At query-time, we simply compute $P(Q|C_j)$ as the cosine similarity between concept vector $C_j$ and the query vector $Q$.

### 3.1.4 Image Relevance for Query $P(I|C_jQ)$.

Finally, we need to compute $P(I|C_jQ)$ to score actual images within the selected concept. Here, we want to favor images contained in communities, as they are more likely to be relevant. However, we need not limit our method to only images from communities. To ensure good coverage on the images, we consider both, images shared within the member communities of the concept, as well as other relevant images that are not members of a community, but are similar to concept vector $C_j$. Thus, we compute $P(I|C_jQ)$ as

$$P(I|C_jQ) = \lambda * membership(I, C_j) + (1-\lambda)CosineSim(I, C_j) \quad (4)$$

where, $membership(I, C_j) = 1.0$, if image $I$ is a member of some community contained in concept $C_j$, and is 0.0 otherwise, $CosineSim(I, C_j)$ is the cosine similarity between image vector $I$ and concept vector $C_j$, and $\lambda$ is the factor used to favor images that are members of a community contained in concept $C_j$. While sophisticated techniqies of computing $\lambda$ can be employed, we observe that setting $\lambda$ at 0.5 was a reasonable way to balance community membership and image similarity. In this way, images that were both, contained in communities as well as similar to concept vector $C_j$ received the highest scores and ranked among the top results.

### 3.1.5 Community Selection Threshold

Since we use communities as concept filters for accessing images, queries with fewer or less relevant communities can suffer in a purely community-based approach. To alleviate this problem of purely community-based approach, we introduce a simple community selection threshold $\alpha$, in which non-community-based images are permitted to appear among the top image search results. That is, out of the top $N$ results that we show to the user, we reserve a fraction $\alpha$ for community-based results and the remaining $(1-\alpha)$ default to plain image search results without the intervention of communities. We vary $\alpha$ depending on the availability of good quality relevant communities for a query. If a query has ample relevant and high quality communities available, we increase $\alpha$ to include more image results from communities, and vice versa. By making sure that $\alpha$ is smaller than one, we are "reserving" the room in the search results for the images that do not belong to one of the matching communities. Note that $\alpha$ does not impact the performance of community-based concept-matched search; it simply improves user's search experience by defaulting to plain search when no results are available from community-based concept-driven search. In Section 4, for true evaluation of community-based approach, we set $\alpha$ to its strictest value of 1.0, i.e., we truly test the pure community-based concept-driven search.

### 3.1.6 Discussion: Communities as Concepts

While the notion of treating communities as concepts is useful, the most popular community may or may not correspond to the most popular query concept that the searcher may be looking for. For example, for the query 'apple', the most popular concept in the searcher's mind is likely to be 'apple, the fruit', while the most popular community may be that of 'Apple, the brand'. However, during our preliminary investigation we observe that, even if the *most* popular concept for searcher and communities may be different, the *top few* concepts from communities most often cover the main topics that the searcher may be looking for. This is further verified by our query-concept evaluation in Section 4.2.

## 3.2 Cluster-Based Concept Extraction

Instead of using communities, concept-driven search can be implemented using keyword clustering, treating clusters generated based on the tags associated with images as concepts. [5] More precisely, given a query, we first select the images with matching query tags. We then cluster those matching images based the tags associated with them using a standard clustering algorithm, such as Latent Semantic Indexing(LSI) [5], Latent Dirichlet Allocation [3] (LDA), or K-means clustering. Once the set of clusters are generated by the algorithm, we consider each output cluster as a possible concept of the given query, hoping that the clustering method would have placed the images from the same concept into the same cluster. Depending on what clustering algorithm is used, one image may belong to multiple clusters and thus may have multiple concepts (e.g., Latent Semantic Indexing [5] and Latent Dirichlet Allocation [3] , both allow an image to be assigned to multiple clusters with different strength in principle).

Once the clusters are generated, we rank the images by considering the popularity of a concept, $P(C_j)$ in equation (3), as the number of images belonging to the cluster. Also, we may represent each cluster by the set of keywords that are most frequently associated with it, so other factors in equation (3), $P(Q|C_j)$ and $P(I|C_jQ)$, can be measured similarly as before under this representation. Here we refrain from considering images from outside the cluster, as these clusters are generated directly from the images themselves, with no image-coverage issues unlike community based approach.

---

[5]Note that the clustering in communitiy based approach is minimal, and is limited to merging only very highly similar communities, whereas in case of cluster based approach, much work is done by the clustering algorithm.

Later, in our experiment section, we employ LSI for cluster-based concept extraction, due to its suitability for clustering tags, its computational performance that allows real-time clustering of matching images, and the wide availability of standard LSI packages.

## 4  Evaluation

Our first goal is to compare concept-driven search with plain search in terms of how well each image search technique performs. Further, among the methods for implementing concept extraction (namely, cluster based and community based), we compare how well each method captures query to concept relevance and concept to image relevance.

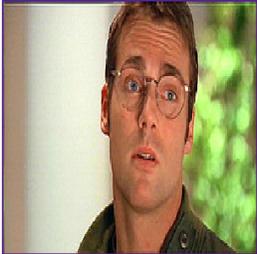

Figure 2: Query to Image Match or Overall evaluation website.

**Query-Concept Match**

**Instructions:**

**Concept:** When a user is searching for a query, he often have some preconceived ideas about it. For example, the query **"jasmine"**, the different concepts are
Jasmine **flower** or
**Princess** Jasmine from Disney or
Jasmine as a **girl's name** or
Jasmine as a **pet's name** or
Jasmine as a type of **Tea**.

In the following evaluation, for each query on the left, we display related concepts represented as a set of keywords.
Please check the radio button 'Good', if you find the suggested concept relevant to the query. When a user enters the query on the left, do you think he means the suggested concept?

**Good:** This concept is relevant to the query for image search.
**Bad:** This concept is not relevant to the query for image search.
**Unclear:** The query or concept is unclear. Cannot say good or bad.

| Query | Concept | Good | Bad | Unclear | Comments |
|---|---|---|---|---|---|
| art | art Collage Vintage | ○ | ○ | ○ | |
| aston martin | aston martin Porsche | ○ | ○ | ○ | |
| beijing | beijing Beijing China | ○ | ○ | ○ | |
| birthday | birthday Birthday Club | ○ | ○ | ○ | |
| boxer | boxer Dog Dog | ○ | ○ | ○ | |
| boxer | boxer Brindle Dog | ○ | ○ | ○ | |
| canada | canada Beautiful Canada | ○ | ○ | ○ | |
| canon | canon Canon EF is USM | ○ | ○ | ○ | |

Figure 3: Concept Extraction Evaluation: Query to Concept Match evaluation website.

## 4.1 Experimental Setup

We implemented tag-based image search and concept-driven search with concept extraction using community-based and cluster-based approaches for a set of crawled Flickr images.

### 4.1.1 Datasets and Tools

To obtain social web data, we crawled Flickr website using their open API service. All data was collected using Perl and XML parser on a Debian Linux machine. We crawled 175,302 images from 44,758 unique users 65,932 communities or Flickr Pools. Flickr identifies about 100 images each day as "interesting". Our crawl has 100% coverage on the interesting photos identified by Flickr between 2004 to 2007, and 100% coverage on the communities in which these photos were shared. Table 5 summarizes the dataset we collected.

For collecting image queries, we built a simple web interface using Flickr data and recruited users to issue queries to the interface as they would normally do on a regular image search engine. From this, we were able to obtain 136 unique queries from 18 unique users in 3 days, which were used as our evaluation data collection.

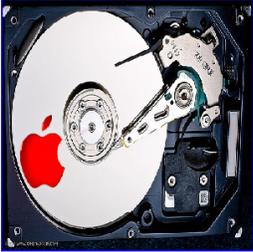

Figure 4: Concept Extraction Evaluation: Concept to Image Match evalution website.

To implement plain search, we used Apache Lucene API [6]. We built an inverted index over the crawled images storing title, description, url, tags, and comments for each image. Similarly, for community-based approach we indexed the crawled communities storing frequent tags with normalized frequency counts, and other associated information like title, description, url, and number of members for each community. We built a search tool over these indexes using scoring functions as described in the Section 3. To implement clustering over the indexed images, we used the framework from Carrot Clustering API [7]. To truly evaluate community-based approach independently, we set the community selection threshold $\alpha$ to its strictest value of 1.0.

---

[6] http://lucene.apache.org
[7] http://project.carrot2.org

| Total Number of unique Users | 44,758 |
| --- | --- |
| Total Number of Images | 175,302 |
| Total Number of Communities | 65,952 |
| Coverage on Flickr 'interesting' photos | 100% upto Dec 2007 |
| Coverage on communities sharing Flickr 'interesting' photos | 100% upto Dec 2007 |

Figure 5: Summary of data collected from Flickr

### 4.1.2 Evaluation Setup

We conducted user studies for all three types of experiments: (1) overall search quality (2) query-concept match and (3) concept-image match. We recruited 27 users for this. These were all anonymous users, identified by their ip address. Each user participated in ALL three types of evaluations. Since we asked each user to evaluate roughly 10 queries or concepts for each type of evaluation and because we wanted to get evaluation on all 136 queries, each query was evaluated by one or (at most) two users.

Figures 4, 4 and 3 show the interfaces of the websites recording user feedback. For query-concept evaluation, for each query we presented top 5 concepts to users and asked them whether the concept looked relevant. For concept-image evaluation, we presented top 5 images from each concept and asked for relevance. For overall evaluation, we presented users with query-image pairs that were output by plain search and the two concept-driven methods, and asked them whether the pair looked relevant. We asked evaluators to rate a pair as 'Good' if they found the pair relevant, 'Bad' - if users found it irrelevant, and 'Unclear' if users were unable to determine the relevance due to ambiguity. The option of 'Unclear' was important to have as all the concepts were machine generated and were presented to the user as a set of keywords. We also collected user comments for each query concept pair. Users mostly wrote comments only for 'Unclear' ratings to explain why the concept was unclear. As our evaluators were not the exact same set of people who issued the queries (though there was significant overlap), we asked the evaluators to provide their ratings based on what *they expect* to see as image search results for the displayed query. Evaluators were allowed to skip result pairs leaving them 'Unrated' in case they did not understand the query. In all experiments, we had < 1% unrated pairs and these were not used for evaluation. In all evaluations, the users were not aware of how the query-image, query-concept, and concept-image pairs were generated. We used this user evaluation data to compare the the methods described in our paper, without manually filtering/choosing any preferred queries (except when there was user disagreement as we explain shortly).

*Inter-User Agreement.* Unfortunately only a small fraction of 136 queries were evaluated by more than one user because we wanted to collect evaluation data for all 136 queries from just 27 users. For the small number (roughly 7%) of evaluation pairs for which we had multiple user inputs, the users did agree on their relevance judgment. Roughly 85% of the pairs, the users indicated the same relevance judgment. We removed the pairs with different user opinions for our evaluation, i.e., we discarded about 1% of all relevance judgements.

## 4.2 Results

We split the evaluation into three parts - query to image match indicating overall search quality, query to concept match, and concept to image match.

We measure the usefulness of concept-driven image search by comparing the overall search quality for the two concept-driven techniques versus plain search. Figure 6(a) shows the average precision at rank $k$ for the top 50 results from each technique. To make the results more readable, the scale of this plot starts at 0.5 average precision value on the Y-axis. Both the concept-driven concept extraction techniques, i.e., cluster-based and community-based, perform much better than plain search. At rank 10, cluster-based concept-driven search outperforms plain search by about 6.67%; community-based concept-driven search outperforms plain search by 20.80%, and cluster-based concept-driven search by 13.47%. At rank 50, community-based approach is 27.45% better than plain search, and 14.02% better than cluster-based approach. Overall, we observe that concept-driven search and specifically community- based concept extraction performed consistently better than both other techniques with the difference being more pronounced with increase in rank. We ran t-test and computed p-values for the precision differences between the methods at top-k results. In all cases, p-values were in the order of 0.0001, indicating that the observed improvement is statistically significant with more than 99% confidence.

After comparing concept-driven search to plain search, we further investigate the two concept-driven search techniques. To measure searcher's perception of goodness and to verify our assumptions that searchers associate keywords with certain concepts, we evaluate query to concept matching for concept-driven search. For each query, we generate relevant concepts using cluster-based and community-based concept extraction, representing each concept by the set of the most frequent keywords in the cluster or community. We present each query-concept pair to our evaluators as shown in Figure 3, and ask them to rate the concepts as relevant or irrelevant to the given query. As seen in figure 6(b), for community-based approach, 64.6% query concept matches as 'Good', 28.3% as 'Bad', and 7.1% as 'Unclear' and for keyword cluster-based approach, 53.5% query concept matches were rated as 'Good', 25.6% as 'Bad', and 20.9% as 'Unclear'. Community-based approach received 20.7% higher rating on 'Good' than cluster-based system in identifying relevant concepts. Unlike community-based system with 7% 'Unclear' ratings, cluster-based approach received over 20% query concept matches as 'Unclear'. The p-values were for t-test for these experiments were in the order of 0.0001. These results indicates that users were unable to identify the concept generated by co-occurrence based clustering of terms. In general, concepts generated from community-based concept extraction were more intuitive to users. This is in line with our hypothesis, that since communities are organized by human users, their concepts are more human interpretable.

Community-based approach picks representative images for each concept by plain TF-IDF based tag search over the communities' shared pool of images. Cluster-based approach picks representative images for each concept by a similar search on each cluster. To verify if image concepts are correctly captured by communities and clusters, we evaluate concept to image match quality of each method. As seen in the plots of Figure 6(c), in case of community-based approach, 70.9% concept to image matches were rated as 'Good', 24.6% as 'Bad', and 4.5% as 'Unclear'. For cluster-based approach, 47.7% concept to image matches were rated as 'Good', 27.3% as 'Bad', and 25.0% as 'Unclear'. Here again, the ambiguity for keyword cluster-based matches is higher indicated by a much larger 'Unclear' rating of 25.0% as against 4.5% for community-based approach. Also, the p-values were for t-test for these experiments were in the order of 0.0001. indicating that the observed result is statistically significant. We observe that community-based approach is at least 23% better than cluster-based approach in 'Good' rating, indicating relevant images.

### 4.2.1 Failure Analysis

We briefly go over what each technique did well and where it failed. While some results of plain search were good, many suffered from irrelevance due to lack of clean tags for images. If some form of tag cleaning and tag weighting is performed over individual images, the results of plain search can improve.

One of the main drawbacks of cluster-based concept extration was its ambiguity in generated concepts. This caused a major drop in its relevance score in every evaluation. Another problem we observed was that the granularity of clusters was at times too fine and at other too coarse. This could be a possible another reason for its weaker performance.

Though community-based approach was the most successful among the three, it was not free from problems. One problem we observed was that image results from some communities were high in relevance, but not good quality images. Evaluators found these images less appealing and rated them as 'Bad'. Though image quality was not our key objective, we learn that query independent quality measure for images as well as communities is necessary for overall success of a technique.

### 4.2.2 Coverage in Communities

In this section, we measure the coverage of communities in our Flickr dataset to measure how many images are favored due to relevant community membership, and how many queries are benefitted by community based approach.

First, we measure how many of our crawled "interesting" images [8] are shared in communities, and stand a chance to benefit from community membership. Figure 7(a) shows the histogram indicating number of communities on the X-axis and the number of images (from our crawled dataset of 'interesting' images) on the Y-axis. About 13.7% of all "interesting" images had zero communities, i.e., these images were not shared in any community. All other images were shared in at least one community with the highest number of communitites for a single image being over 300. Thus, the interesting photos on Flickr are reasonably well shared with about 86% images having 1 or more communities. This number, though encouraging in the 'interesting' image set, is possibly lower if we consider all images in the Flickr database.

---

[8]Flickr marks certain images as "interesting".

Second, we measure the coverage of communities in our query log. Figure 7(b) shows the histogram indicating the number of queries on the Y-axis and the number of matching communities on the X-axis. The histogram bin sizes on X-axis are chosen to highlight the more interesting categories in the distribution. We observe that most queries had a over 100 matching communities available. Only 9 out of 136 input queries found no communities. Further, 36 queries mapped to communities with very low score for popularity ($< 10$ users) and were discarded. Thus, only 91 of 136 queries were truly answerable using community-based approach, i.e., only 66.9% queries could find relevant communitites and make use of them to improve search results. For the other 33.1% queries with no matching communities, and could not benefit from community based concept extration.

## 4.3 Image Results Interface

Since the concept-driven search explicitly identifies the set of relevant concepts given a query, it naturally leads to the cluster-oriented result page, where each cluster represents a concept. That is, as shown in Figure 8(a), for the concept-driven interface we may explicitly display concepts in the result (together with the representative labels for the concept), ranked by their relevance and popularity, and showing a few sample images for each concept. In our informal survey of our users, we find that users perceive image clustered interface more appealing for search results than ranked image lists as long as the identified clusters are meaningful and easy to understand.

However, to make sure that our evaluation is comparable to plain search results that does not employ concept-based clustering idea, we also implemented a ranked list interface and used for our evaluation, as shown in Figure 8(b). In this, we pick images from all concepts and score them by the combined concept score and individual image relevance score and generate a simple ranked list of images.

*Sample Results* - From our evaluation, we noticed that some of the concepts identified by our community based method often captures intuitively meaningful cluster of images surprisingly well. To give a taste of our results, we refer again to Figure 8, which displays the actual output of community-based concept matching. In the figure, we show the top-3 concepts that our method identifies together with the high frequency tags associated with those concepts. The first concept that we show is about "orange color" and we can see that the top-2 tags indeed are orange and color. The second concept is about "orange fruit". The keyword "fruit" is not one of the top-3 tags, but interestingly we see that the keyword orange appears as plural, indicating that it is more likely refering to fruits, and foreign language terms for orange, "naranja" and "arance", are part of the top-3 tags for this concept. The third concept is about "orange-colored flower" and the tags correctly communicate the meaning of this concept.

### 4.4 Applicability of Proposed Methods

For online content sharing websites, one of their primary goals is to share data with others of similar interest. We observe that many "Web 2.0" sites have active user communities. For example, social websites supporting media contents, like Flickr [9], YouTube [10], PhotoBucket [11], Daily Motion [12], Smugmug [13], MySpace [14], and Facebook [15], all allow users to form communities and interest groups. The applicability of our methods depends on the availability and the type of contents that are shared through communities. For example, we believe Youtube and Smugmug are excellent candidates for our methods, because user communities are actively used to share and recommend contents to others. On the other hand, while Facebook has a large number of communities, the website only recently allowed users to generate and share data through communities. So there is not much media content associated with communities yet, but with increasing participation of users, we can hope that community-based content on Facebook also increases.

## 5 Related Work

We discuss three lines of research related to our work: image search, tag-based search, and language model-based and topic-based document matching approaches to text search.

Image search research has explored the use of image content or visual information, link information, and automatically generated or manually provided textual descriptions for retrieving relevant images. Visual or content-based retrieval methods [17, 15] for image search, though useful, suffer from gaps between low-level visual descriptions and a user's semantic expectation [15, 4]. Approaches exploiting link information for images [4] are also less effective for social media content, as these pages are very weakly linked. Text-based image retrieval methods [6, 14, 20] have traditionally relied on texual description of images, generated either automatically from the webpage embedding the image, or by manual annotation [18, 16]. Despite the availability of reasonably large amount of text descriptions on web pages, text-based image retrieval methods face challenges of inconsistencies between user textual queries and image annotations [20].

---

[9]http://www.flickr.com
[10]http://www.youtube.com
[11]http://www.photobucket.com
[12]http://www.dailymotion.com
[13]http://www.smugmug.com
[14]http://www.myspace.com
[15]http://www.facebook.com

The problem of inconsistent terminologies is exacerbated in social media sharing websites, like Flickr, YouTube, using tag-based retrieval, where the length of text associated with user-created images and videos tends to be significantly short. Also, image tags often cover a broad spectrum of the semantic space [16], like where the photo was taken [11], who or what is on the photo, and when the photo was taken. The language used to tag user-created rich media is often very informal and inconsistent. For instance, many users on Flicker tag the image of their cars as "My baby", meaning that the cars are dear to their heart, not that they are baby pictures. These factors make traditional TF-IDF like measures less effective for capturing relevance.

To make tags more usable for search, previous works corroborate the importance of use of additional information in tag-based retrieval. In [1], the authors describe experiments in which annotations in social media, when interpreted without background knowledge actually worsen the understanding of the meaning. Researchers have explored the idea of augmenting tags with additional information to improve relevance of search on these individual objects, either by using classification-based [6, 10] approaches, or using clustering-based [4] approaches. Among classification-based approaches [7, 12, 9] for tag-based retrieval, the general idea is to induce taxonomies over individual object tags. Researchers have explored use of clustering [19, 2] social media contents to generate more context. [2] studies the use of automated tag clusters for better search and browsing on social content. Flickr too provides a mechanism for exploring tags and associated images as a clustered interface [16].

In line with our concept-driven model, previous works have explored the idea of topic-based document retrieval [5, 3] by using probabilistic models to simulate document generation. The Latent Dirichlet Allocation (LDA) model [3] is based on the idea that each document is a mixture of a small number of topics and that each word's creation is attributable to one of the document's topics. Others have demonstrated that cluster-based language models [13] can be more effective than simple document-based retrieval. While our approach is similar to existing clustering-based retrieval work, one very important difference is that we not only consider the query-generation likelihood from cluster, but also the overall "popularity" of the cluster/cencept. The consideration of popularity is crucial for the high-quality results from our methods, because it allows us to return what the users "expect".

# 6 Conclusion

In this paper, we explored the idea of concept-driven search for social media content. When the text associated with media content is limited, searching is difficult. The difficulty in capturing relevance of the content for a searcher is due to the mismatch in expected concepts of searcher and content creator. We observed that concept-driven search along with the notion of "popularity" of the concept is effective in solving the relevance problems in social media search. We provided a probabilitic model to derive ranking metrics for concept-driven search. We implemented and evaluated the use of cluster-based approach and a novel community-based approach for implementing concept-driven search.

---

[16]Examples of Flickr clusters: http://flickr.com/photos/tags/jasmine/clusters

Our results show that concept-driven search clearly outperforms plain search in overall search quality. Among the two techniques for concept-driven search, though community-based approach suffered from query coverage issues, it was generally more successful in capturing the intended meaning of a tag from a content creator and a keyword from a searcher. The primary drawback of cluster-based approach was that the clusters, built merely by keyword co-occurrence relationships, did not generate human interpretable or appealing concepts. Communities, on the other hand, acting as social clusters, naturally incorporated the notion of popularity and implicitly allowed selective aggregation of opinions of a relevant group of users. Compared to clustering, the concepts generated from communities were generally found to be more intuitive.

Social media content in general is ad hoc and not well organized. We believe that our results demonstrate that the apparently chaotic and unstructured contents on social web sites can be made more meaningful if some auxiliary information provided by the users, such as user-created communities, is carefully analyzed and exploited.

# References

[1] J. Bar-Ilan, S. Shoham, A. Idan, Y. Miller, and A. Shachak. Structured vs. unstructured tagging: A case study. In *Proc. of the Collaborative Web Tagging Workshop (WWW 2006)*, May 2006.

[2] G. Begelman, P. Keller, and F. Smadja. Automated tag clustering: Improving search and exploration in the tag space. 2006.

[3] D. M. Blei, A. Y. Ng, and M. I. Jordan. Latent dirichlet allocation. *Journal of Machine Learning Research*, 3:993–1022, 2003.

[4] D. Cai, X. He, Z. Li, W.-Y. Ma, and J.-R. Wen. Hierarchical clustering of www image search results using visual, textual and link information. In *MULTIMEDIA '04: Proceedings of the 12th annual ACM international conference on Multimedia*, pages 952–959, New York, NY, USA, 2004. ACM.

[5] S. Deerwester, S. T. Dumais, G. W. Furnas, T. K. Landauer, and R. Harshman. Indexing by latent semantic analysis. *Journal of the American Society for Information Science*, 41(6):391–407, 1990.

[6] C. Frankel, M. J. Swain, and V. Athitsos. Webseer: An image search engine for the world wide web. Technical report, Chicago, IL, USA, 1996.

[7] S. Golder and B. A. Huberman. The structure of collaborative tagging systems, Aug 2005.

[8] M. A. Hearst and J. O. Pedersen. Reexamining the cluster hypothesis: scatter/gather on retrieval results. In *SIGIR '96: Proceedings of the 19th annual international ACM SIGIR conference on Research and*

*development in information retrieval*, pages 76–84, New York, NY, USA, 1996. ACM.

[9] P. Heymann and H. Garcia-Molina. Collaborative creation of communal hierarchical taxonomies in social tagging systems. Technical Report 2006-10, Stanford University, April 2006.

[10] J. Hu and A. Bagga. Categorizing images in web documents. *IEEE MultiMedia*, 11(1):22–30, 2004.

[11] L. S. Kennedy and M. Naaman. Generating diverse and representative image search results for landmarks. In *WWW '08: Proceeding of the 17th international conference on World Wide Web*, pages 297–306, New York, NY, USA, 2008. ACM.

[12] R. Li, S. Bao, Y. Yu, B. Fei, and Z. Su. Towards effective browsing of large scale social annotations. In *WWW '07: Proceedings of the 16th international conference on World Wide Web*, pages 943–952, New York, NY, USA, 2007. ACM.

[13] X. Liu and W. B. Croft. Cluster-based retrieval using language models. In *SIGIR '04: Proceedings of the 27th annual international ACM SIGIR conference on Research and development in information retrieval*, pages 186–193, New York, NY, USA, 2004. ACM.

[14] Y. Liu, T. Qin, T.-Y. Liu, L. Zhang, and W.-Y. Ma. Similarity space projection for web image search and annotation. In *MIR '05: Proceedings of the 7th ACM SIGMM international workshop on Multimedia information retrieval*, pages 49–56, New York, NY, USA, 2005. ACM.

[15] S. McDonald and J. Tait. Search strategies in content-based image retrieval. In *SIGIR '03: Proceedings of the 26th annual international ACM SIGIR conference on Research and development in informaion retrieval*, pages 80–87, New York, NY, USA, 2003. ACM.

[16] B. Sigurbjörnsson and R. van Zwol. Flickr tag recommendation based on collective knowledge. In *WWW '08: Proceeding of the 17th international conference on World Wide Web*, pages 327–336, New York, NY, USA, 2008. ACM.

[17] J. R. Smith and S.-F. Chang. Visually searching the web for content. *IEEE MultiMedia*, 4(3):12–20, 1997.

[18] L. von Ahn and L. Dabbish. Labeling images with a computer game. In *CHI '04: Proceedings of the SIGCHI conference on Human factors in computing systems*, pages 319–326, New York, NY, USA, 2004. ACM.

[19] X.-J. Wang, W.-Y. Ma, Q.-C. He, and X. Li. Grouping web image search result. In *MULTIMEDIA '04: Proceedings of the 12th annual ACM international conference on Multimedia*, pages 436–439, New York, NY, USA, 2004. ACM.

[20] C. Zhang, J. Y. Chai, and R. Jin. User term feedback in interactive text-based image retrieval. In *SIGIR '05: Proceedings of the 28th annual international ACM SIGIR conference on Research and development in information retrieval*, pages 51–58, New York, NY, USA, 2005. ACM.

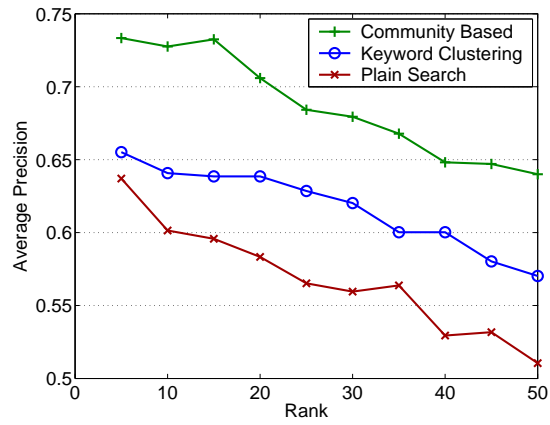
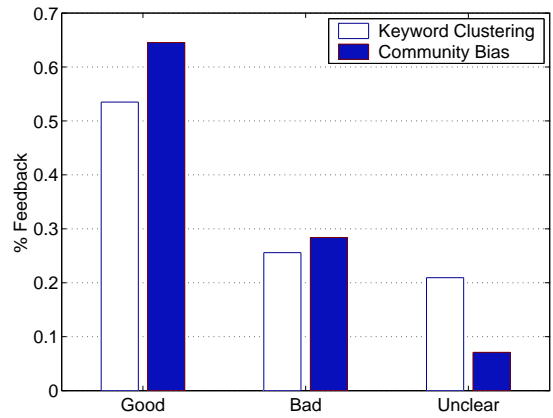
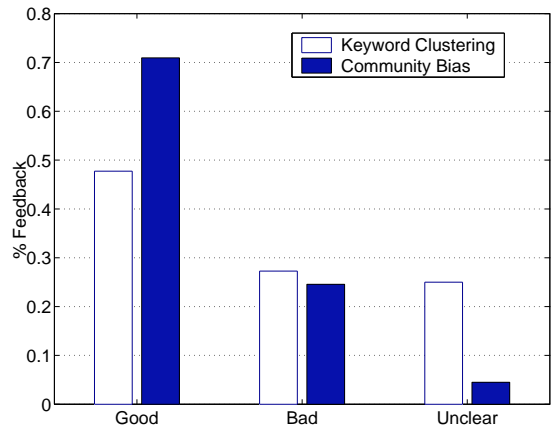

Figure 6: (a) Average Precision at Rank $k$ for the three techniques. (b) Concept Extraction Evaluation: Query to Concept Match. (c) Concept Extraction Evaluation: Concept to Image Match.

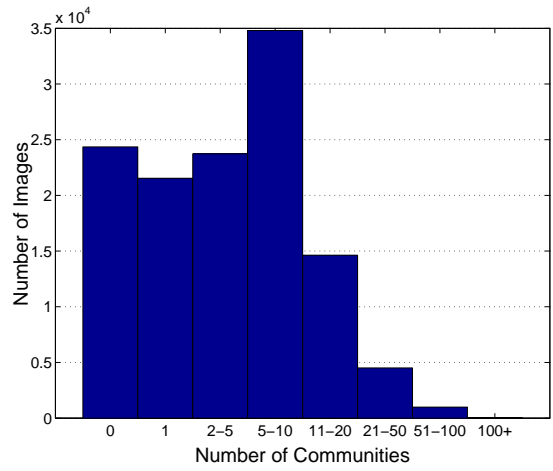

(a)

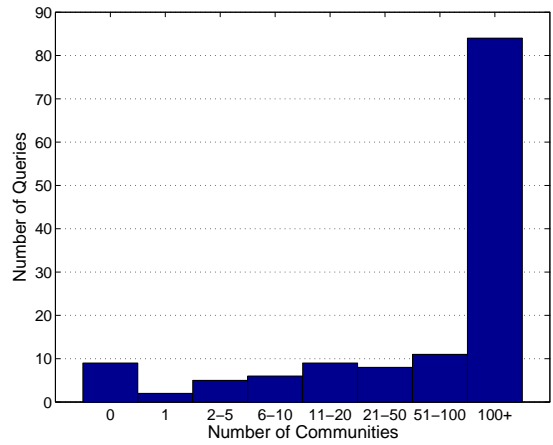

(b)

Figure 7: (a): Histogram for number of communities per image. (b): Histogram for number of communities per query.

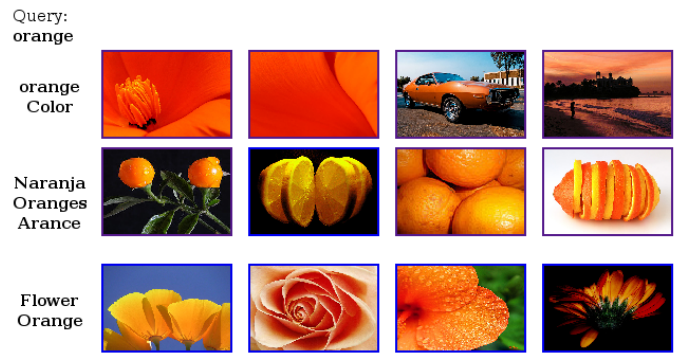

(a)

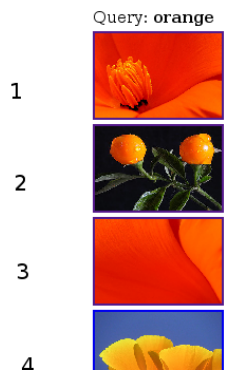

(b)

Figure 8: (a): Results interface as clusters for query 'orange'. (b): Results interface as a ranked list for query 'orange'.